# Ionic Conductivity of a Lithium-Doped Deep Eutectic Solvent: Glass Formation and Rotation-Translation Coupling

A. Schulz, P. Lunkenheimer*, and A. Loidl

* Corresponding author: peter.lunkenheimer@physik.uni-augsburg.de

Affiliation:
Experimental Physics V, Center for Electronic Correlations and Magnetism, University of Augsburg, 86135 Augsburg, Germany

**ABSTRACT:**

Deep eutectic solvents with admixed lithium salts are considered as electrolytes in electrochemical devices like batteries or supercapacitors. Compared to eutectic mixtures of hydrogen-bond donors and lithium salts, their raw-material costs are significantly lower. Not much is known about the glassy freezing and rotation-translation coupling of such systems. Here we investigate these phenomena by applying dielectric spectroscopy to the widely studied deep eutectic solvent glyceline, to which 1 and 5 mol% LiCl were added. Our studies cover a wide temperature range, including the deeply supercooled state. The temperature dependences of the detected dipolar reorientation dynamics and of the ionic dc conductivity reveal the signatures of glassy freezing. In comparison to pure glyceline, lithium admixture leads to a reduction of ionic conductivity, which is accompanied by a slowing down of the rotational dipolar motions. However, this reduction is much smaller than for DESs, where one main component is a lithium salt, which we trace back to the smaller glass-transition temperatures of the lithium-doped DESs. In contrast to pure glyceline, the ionic and dipolar dynamics become increasingly decoupled at low temperatures and obey a fractional Debye-Stokes-Einstein relation as previously found in other glass-forming liquids. The obtained results demonstrate the relevance of decoupling effects and of the glass transition for the enhancement of the technically relevant ionic conductivity in such lithium-doped DESs.

## 1. INTRODUCTION

In deep eutectic solvents (DESs) the mixing of two or more components leads to a melting-point reduction and a liquid state around room temperature. Typically, they are composed of a molecular hydrogen-bond donor (HBD) and a salt representing a hydrogen-bond acceptor (HBA). Many of their properties like sustainability, ease of production, and reduced flammability are superior to those of other solvents used, e.g., in material synthesis or employed as electrolytes in electrochemical devices.[1,2,3,4,5,6,7,8,9,10]

For the latter, DESs containing suitable ions, as Li$^+$ for lithium-ion batteries, are of special interest. Possible candidates are DESs produced by mixing a HBD with a lithium salt.[11,12,13,14,15,16] However, the required eutectic composition ratio implies the use of a large fraction of lithium salts for their synthesis, exceeding that in conventional electrolytes and giving rise to high production costs.[17] An alternative approach is the addition of relatively small amounts (several mol%) of a lithium salt to a low-cost DES.[17,18,19,20,21,22,23] Among them, so-called natural DESs (NADESs), whose constituents are found in nature, are an interesting option due to their environmental friendliness, non-toxicity, and renewability. In the present work, we employ the widely investigated NADES glyceline, a 1:2 mixture of choline chloride (ChCl) and glycerol, as a lithium-salt solvent. Its organic-salt component has many benefits as it is relatively inexpensive and biodegradable.[10] Moreover, ChCl-based DESs were found to reveal significantly higher ionic conductivity than corresponding Li-salt-based DESs with the same HBD.[16,24] High conductivity is required for the application of DESs as electrolytes in electrochemical devices.

As often found for eutectic mixtures, many DESs exhibit a glass transition at low temperatures and the concomitant characteristic properties of supercooled liquids.[3,11,16,25,26,27] As pointed out in several earlier works from our group, this can also affect their room-temperature properties, including the technically relevant dc conductivity.[16,26,28,29,30] Moreover, most DESs contain dipolar entities (e.g., the glycerol molecules in glyceline), which can reveal reorientational dynamics that is intimately connected with the glassy freezing detected in the viscosity.[28] Such dipole rotations in ionic conductors are also of interest because they can play an important role in the enhancement of ionic mobility as was explicitly shown for ionic liquids and various crystalline systems.[31,32,33,34] Interestingly, varying degrees of correlations of reorientational dipolar and translational ionic motions were recently found for a number of DESs, too.[16,26,27,28,29,35,36] However, overall the glassy freezing and dipolar dynamics of DESs were only rarely investigated until now. To our knowledge, only in Ref. 18 the glass transition in a lithium-doped DESs was briefly discussed, based on temperature-dependent conductivity and differential scanning calorimetry (DSC) measurements.

To investigate these phenomena, dielectric spectroscopy is a well-established experimental method. It can simultaneously provide information on both the dipole and the ion dynamics.[37,38] Here, we apply this method to glyceline with the admixture of 1 or 5 mol% LiCl. Adding LiCl avoids the introduction of another anion species, which even may have a dipolar moment as in the often-used Lithium

bis(trifluoromethanesulfonyl)imide (LiTFSI). This should ease the interpretation of the obtained results.

## 2. METHODS

Glycerol was purchased from Sigma-Aldrich, ChCl from Alfa Aesar (both, 99% purity), and LiCl from Merck (98%+ purity) and used as is. Both samples were prepared by admixing the respective amounts of LiCl (1 and 5 mol%) to the eutectic composition between ChCl and glycerol (1:2 molar ratio) inside a glass vessel. The mixture was stirred at 328 K until a homogeneous and clear liquid without any crystalline residue was attained. Since water in DESs has a strong effect on their intrinsic properties,[3,27,39,40] the samples were kept in a dry nitrogen atmosphere during the measurements. Additionally, the water content of both samples was determined via coulometric Karl-Fischer titration to be less than 0.05 wt.%, deeming its concentration negligible.

The dielectric measurements were conducted with a frequency-response analyzer (Novocontrol Alpha-A analyzer) covering a broad frequency range between 0.1 Hz and 10 MHz. For this purpose, the samples were put into a stainless steel parallel-plate capacitor with a diameter of 12 mm. The plate distance was kept at 0.1 mm using glass-fiber spacers. To study the dipolar and ionic dynamics in a broad temperature window, the temperature of the sample cell was regulated by a closed-cycle refrigerator. In addition, DSC measurements were conducted using a DSC 8500 from PerkinElmer, applying heating and cooling rates of 10 K/min.

## 3. RESULTS AND DISCUSSION

**3.1. Dielectric spectra.** Dielectric measurements of ionically conducting materials can be analyzed in terms of the complex dielectric permittivity, the complex conductivity, or the dielectric modulus.[37,41] The present DESs are ionic conductors with additional reorientational degrees of freedom due to the enclosed dipolar entities: the glycerol molecules and the choline cations. It should be noted that in the pioneering work by Macedo et al.,[41] introducing the modulus representation and its conductivity-relaxation interpretation, only "ionic conductors which contain no permanent molecular dipoles" were considered.[41] Indeed, various problems can arise when applying the modulus formalism to systems with simultaneous ionic translational and dipolar reorientational degrees of freedom. This is discussed for DESs in some detail in Refs. 16 and 28. Therefore, here we adhere to the approach in our earlier works on ionic conductors,[33,34,42] including DESs[16,26,28,29,30] analyzing the complex dielectric permittivity, $\varepsilon^* = \varepsilon' - i\varepsilon''$ where $\varepsilon'$ is the dielectric constant and $\varepsilon''$ the dielectric loss. It is commonly used for materials with reorientational degrees of freedom[37,38] and also facilitates a direct comparison with our previous results. In addition, we consider the real part of the conductivity $\sigma'$, enabling the direct identification of the dc conductivity in the spectra. Via $\sigma' = \varepsilon''\varepsilon_0\omega$ (where $\varepsilon_0$ is the permittivity of free space and $\omega = 2\pi\nu$) it is directly related to the loss.

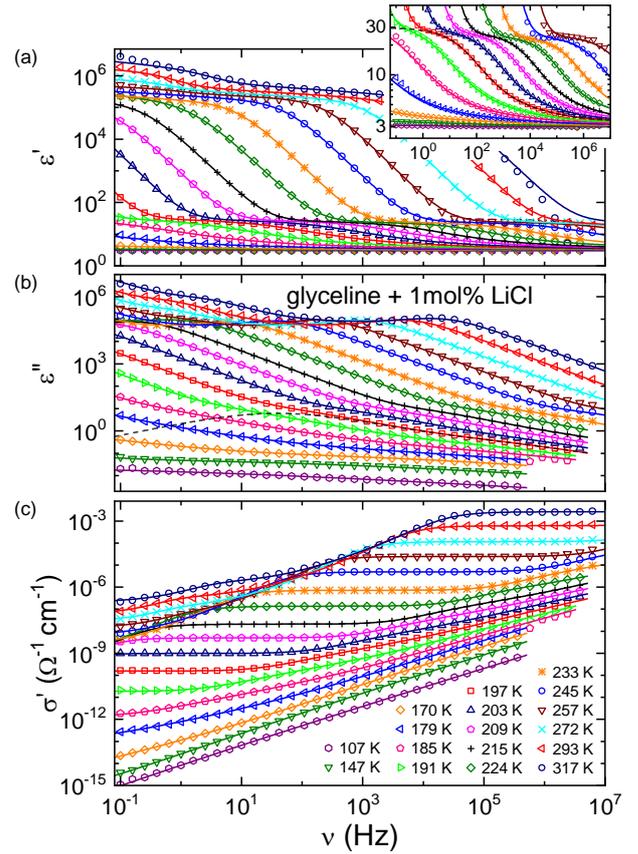

Figure 1. Spectra of the dielectric constant $\varepsilon'$ (a), loss $\varepsilon''$ (b), and conductivity $\sigma'$ (c) of glyceline with 1 mol% LiCl, as measured at various temperatures. The inset provides an enlarged view of the intrinsic relaxation steps in $\varepsilon'(\nu)$. The solid lines in (a) and (b) are fits using an equivalent-circuit approach as described in the text, including a dc-conductivity contribution and two intrinsic relaxation processes ($\alpha$ and secondary). The fits were simultaneously performed for $\varepsilon'$ and $\varepsilon''$ and the fit lines shown for the conductivity were calculated using $\sigma' = \varepsilon''\varepsilon_0\omega$. The dashed lines in (b) and in the inset exemplarily show the intrinsic relaxation contributions for 197 K.

Figure 1 shows spectra of $\varepsilon'$, $\varepsilon''$, and $\sigma'$ as obtained for glyceline with 1 mol% LiCl at various temperatures. The sample with 5 mol% salt content revealed qualitatively similar results. The main features in these spectra are the same as previously found for other ionically conducting DESs:[16,27,28,29,30,35] At high temperature and low frequencies, $\varepsilon'(\nu)$ (Figure 1a) approaches "colossal" values,[43] exceeding $10^5$ in the present case. This is due to electrode polarization, a common finding for ionic conductors.[44,45] It arises from a space-charge region close to the sample surface that forms when the ions reach the electrodes at low frequencies, where their motion essentially becomes blocked. It is well known that electrode polarization leads to a so-called Maxwell-Wagner relaxation, which shows up by a step-like decrease of $\varepsilon'(\nu)$ with increasing frequency as also found for intrinsic dipolar relaxations.[45] It is the dominating feature in the present $\varepsilon'$ spectra. For the highest temperatures, at frequencies below this huge relaxation step, a second one is revealed in Figure 1a. Such behavior is often seen in the spectra of ionic conductors when sufficiently low frequencies and high temperatures are covered.[45] It also can be ascribed to non-intrinsic electrode effects, pointing to a second, slower ion-transport mechanism within the space-charge region.[45,46] Maxwell-Wagner relaxations also lead to peaks or shoulders in the loss spectra, occurring at the same frequency as the points of inflection of the $\varepsilon'$ steps, as indeed seen in Figure 1b. Due to the relation $\sigma' \propto \varepsilon''\nu$, this effect also causes the decrease of $\sigma'(\nu)$ with decreasing frequencies, detected at high temperatures in Figure 1c.

At frequencies beyond the Maxwell-Wagner-dominated regime, another step-like decrease in $\varepsilon'(\nu)$ shows up (see inset of Figure 1 for an enlarged view). In analogy to other DESs,[16,26,27,28,29,30,35,36] we ascribe it to an intrinsic dipolar relaxation process due to reorientational motions of the dipolar entities in this DES, termed $\alpha$ relaxation. With decreasing temperature, these relaxation steps strongly shift to lower frequencies, indicating the continuous slowing down of the molecular dynamics typical for dipolar glass-formers.[38,47] This is confirmed by the occurrence of a typical glass-transition anomaly in the DSC results obtained for this DES (Figure S1). A similar intrinsic dipolar relaxation feature was previously also reported for pure glyceline.[26] There it was assumed to be mainly due to the reorientational motions of the glycerol molecules, based on the fact that at high temperatures the corresponding relaxation times are of similar magnitude as for pure glycerol. Recent nuclear magnetic resonance (NMR) experiments[48] also confirmed that glycerol reorientations play a major role for the dielectrically detected relaxation in glyceline. In that work, the reorientations of the choline ions were found to exhibit a similar, but somewhat faster rotation rate (by about 30% at room temperature). In marked contrast, based on plots of the frequency-dependent derivative of $\varepsilon'$,[49] very recently, the existence of a reorientational choline-ion relaxation process that is slower than the glycerol relaxation was claimed.[36] Whatsoever, the present dielectric results do not reveal any indications of two overlapping relaxation processes. We also checked plots of the derivative of $\varepsilon'$, but did not find any evidence of a second relaxation. In this context, one should note that it is quite a common phenomenon that dielectric spectroscopy only detects a single relaxation process in mixtures comprising two dipole species[50,51] and NMR certainly is better suited to deconvolute dipolar motions occurring on similar time scales.

In principle, dipolar reorientations should lead to peaks in the dielectric loss spectra, shown in Figure 1b. However, in the present case, these peaks are partly superimposed by the strong contribution from the dc conductivity $\varepsilon''_{dc} \propto \sigma_{dc}/\nu$, as is typical for conducting materials. This leads to a $1/\nu$ divergence of $\varepsilon''(\nu)$ towards lower frequencies. Therefore, only the high-frequency flanks of the relaxation peaks are observed, giving rise to a crossover from the $1/\nu$ to a shallower decline of $\varepsilon''(\nu)$. This is illustrated by the dashed line in Figure 1b exemplarily indicating the intrinsic relaxation contribution to the 197 K spectrum.

At sufficiently low temperatures, when the spectral features associated with the $\alpha$-relaxation in Figure 1a and b have shifted out of the frequency window, a region with only weak frequency dependence is revealed, evidencing an additional, minor contribution in this region. It points to the presence of a broad secondary relaxation that is faster than the $\alpha$ relaxation. Such processes, usually termed $\beta$ relaxations, are commonly found in glass-forming liquids[34,52,53,54,55] and also were detected in DESs.[16,25,26,28,29,30] Their detailed discussion is out of the scope of the present work.

The frequency-independent plateau observed in the conductivity spectra of Figure 1c marks the ionic dc conductivity. Its strong many-decades decrease with decreasing temperature evidences the essentially thermally activated ionic charge transport (as treated in the following section, in fact it varies even stronger than expected for canonical thermal activation). At frequencies beyond the dc plateau, $\sigma'(\nu)$ starts to increase with frequency. This can be ascribed to the dipolar-relaxation contributions treated in the above discussion of the loss spectra (Figure 1b), since $\varepsilon''$ and $\sigma'$ are closely related via $\sigma' \propto \varepsilon''\nu$.

It should be noted that qualitatively similar behavior as seen in Figure 1c is also expected within various models for ionic hopping conductivity, which can lead to significant ac conductivity, usually approximated by a power-law increase of $\sigma'(\nu)$.[56,57,58,59] However, such models only treat translational motions of ions and are not intended to account for reorientational dynamics which is inevitable in DESs with dipolar HBDs and/or ions. As we are able to fit the present experimental data without assuming any ac-conductivity contribution (see detailed discussion below), we refrain here from employing such models. If ac conductivity exists in our sample, it is completely superimposed by the high dc conductivity (at low frequencies) and by the spectral features from dipole-reorientations (at high frequencies). Thus, including ac conductivity in the fits would contradict Occam's razor. This issue was discussed in more detail in several previous works from our group, where data on different ionic conductors were also analyzed by alternative fits including contributions from hopping conductivity.[16,26,28,34]

To reliably determine the intrinsic relaxation parameters, it is crucial to fit the dielectric spectra with the inclusion of the non-intrinsic electrode effects discussed above. For this purpose, we model the weakly conducting space-charge region by a distributed parallel RC circuit.[45] Assuming a series connection of this circuit to the bulk part of the sample corresponds to an equivalent-circuit description as previously applied to various other ionic conductors,[33,34,45,60] including DESs.[16,26,28,29,30] To fit the intrinsic $\alpha$ and secondary relaxations we used the sum of two Cole-Cole (CC) functions, an empirical function that is often applied to glass-forming materials and leads to symmetric loss peaks.[37,38,61,62] It is given by:

$$\varepsilon^* = \varepsilon_\infty + \frac{\Delta\varepsilon}{1 + (i\omega\tau)^{1-\alpha}} \qquad (1)$$

Here, $\varepsilon_\infty$ is the high-frequency dielectric constant, $\Delta\varepsilon$ is the relaxation strength, and the exponent $\alpha$ determines the broadening compared to the prediction of the Debye theory, the latter corresponding to eq 1 with $\alpha = 0$. While the $\alpha$ relaxation in glass-forming liquids often (but not always) reveals asymmetric peak shapes, in pure glyceline the CC function was reported to provide good fits of the dielectric spectra.[26,63] In a DES formed by an 1:2 mixture of $ZnCl_2$ and ethylene glycol, the situation is similar.[29] The loss peaks described by the CC function are broader than predicted by the Debye theory of dipolar relaxation, which in glass-forming liquids is commonly ascribed to a distribution of relaxation times due to heterogeneity.[64,65] Between about 180 and 230 K, where the width parameter $\alpha$ in eq 1 can be determined with sufficient precision, it decreases upon heating from about 0.5 to 0.35 for the sample with 1 mol% LiCl and from about 0.6 to 0.4 for 5 mol%. Similar narrowing of the $\alpha$ relaxation with increasing temperature is often found in glass-forming liquids.[47] The stronger broadening found for the 5 mol% sample points to stronger heterogeneity due to the higher amount of admixed salt.

Finally, the contribution of the dc-conductivity to the loss was taken into account by the term $\varepsilon''_{dc} = \sigma_{dc}/(\varepsilon_0 \omega)$. The solid lines in Figure 1 present the fits achieved using this approach, which were conducted simultaneously for $\varepsilon'(\nu)$ and $\varepsilon''(\nu)$ (the fit lines for $\sigma'$ were calculated from those for $\varepsilon''$). They provide a nearly perfect description of the experimental data. One should note that, for each temperature, only part of the different contributions had to be included in the fit function. This is due to the fact that, e.g., at high temperatures the secondary relaxation is shifted out of the frequency window or at low temperatures the electrode effects play no role. This avoids an unreasonably high number of fit parameters.

**3.2. Dc conductivity and $\alpha$-relaxation time.** Figure 2a shows an Arrhenius plot of the temperature-dependent dc conductivities as deduced from the fits of the dielectric spectra of the two investigated samples (closed symbols). The detected room-temperature values are relatively high, reaching technically relevant values of the order of $10^{-3}$ $\Omega^{-1}$cm$^{-1}$. The found nonlinear behavior in this representation evidences clear deviations from the simple thermally-activated charge transport [essentially leading to $\sigma_{dc} \propto \exp(-E/T)$, where $E$ is the energy barrier in K] expected for conventional ionic conductors. Similar non-Arrhenius temperature dependence was also reported for several other DESs that were investigated in a sufficiently broad temperature range.[11,16,18,25,26,27,28,29,30,63] This is typical for ionically-conducting glass-forming liquids[66] and signifies that the ionic mobility in principle is coupled to the non-Arrhenius behavior, characteristic of the structural dynamics of glass-forming liquids and detected, e.g., in their temperature-dependent viscosity.[38,47,67]

As revealed in Figure 2a, the dc conductivity of glyceline with 5 mol% LiCl is lower than that of the sample with 1 mol%, the difference reaching up to a factor of ~4 at low temperatures. At first glance, this seems unexpected because the number density of charge carriers should be higher in the 5 mol% sample and the added small Li$^+$ ions may be expected to significantly contribute to the overall conductivity. However, it is well known that the admixture of lithium ions can reduce the ionic conductivity of liquids as, e.g., previously shown for ionic liquids[42] and also for DESs.[18,23] This trend is also confirmed when comparing $\sigma_{dc}(T)$ of the 1 mol% sample to that of pure glyceline from Ref. 28, shown by the crosses in Figure 2a. This effect can be ascribed to the small size of the Li$^+$ ion which leads to a large charge/ion-radius ratio (sometimes termed "ionic potential") generating enhanced interactions between the ion species in this DES and also between the Li$^+$ cations and the HBD molecules. Indeed, a recent investigation of glyceline with added LiTFSI salt, using fluorescence, ESR, and NMR methods, points to a perturbation of the nanostructural organization triggered by the addition of the lithium salt.[20] Moreover, recent molecular dynamics simulations of glyceline with added LiTFSI salt[21] indicate the formation of clusters between lithium and chloride ions for increasing salt concentration, subsequently leading to a reduction of the ion diffusivity. For the above reasons, lithium addition can be assumed to cause an increase of the viscosity. Such a viscosity increase was indeed experimentally detected for the admixture of LiCl to the DES reline (composed of urea and choline chloride).[19] A viscosity increase due to LiCl admixture should lead to a conductivity reduction, having in mind the somewhat oversimplified picture of spheres (valid at least for the lithium and chloride ions) translationally moving within a viscous medium, whose motions become successively impeded when viscosity increases. This picture is quantified by the Stokes-Einstein relation predicting $D_t \propto T/\eta$, where $D_t$ denotes the translational diffusion coefficient. Using in addition the Nernst-Einstein relation ($\sigma_{dc} \propto D_i/T$ where $D_i$ is the ionic diffusion coefficient), and assuming $D_i = D_t$, gives $\sigma_{dc} \propto 1/\eta$.[68,69,70,71]

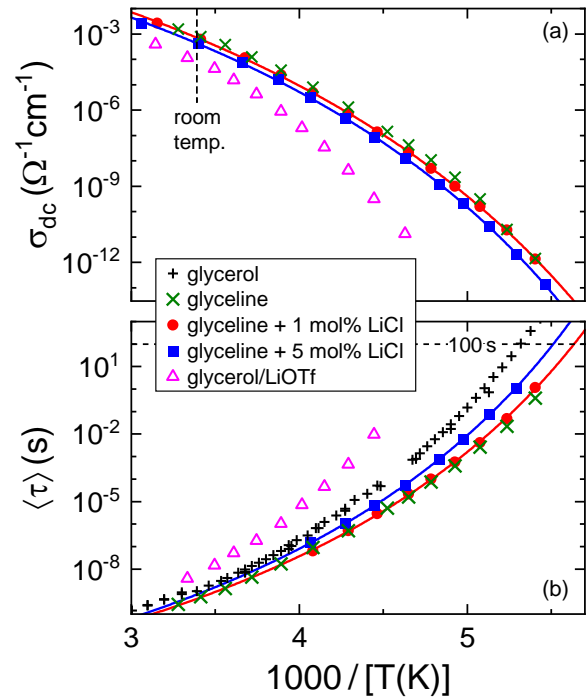

Figure 2. Temperature-dependence of the dc conductivity (a) and mean $\alpha$-relaxation time (b) for glyceline with two concentrations of LiCl (closed symbols) and for pure glyceline taken from Ref. 28 (crosses). For comparison, data from Ref. 16 for a DES formed by glycerol and LiOTf are included (open triangles). In addition, in (b) $\tau(T)$ for pure glycerol is shown (plusses).[78] The lines in (a) and (b) are fits with the VFT laws, eqs 2 and 3, respectively. The vertical dashed line in (a) denotes room temperature (295 K). The horizontal dashed line in (b) indicates $\tau(T_g) = 100$ s.

As an example of a DES produced by the mixing of a HBD with a lithium salt, the latter acting as the HBA, Figure 2a contains data on glycerol mixed with lithium triflate (LiOTf) in a eutectic 3:1 molar ratio (triangles).[16] The much larger lithium content, compared to the present samples, leads to significantly reduced conductivity. This underlines the advantage of Li-addition to a DES like glyceline, i.e., without a lithium salt forming a major constituent, when considering electrolyte applications.

In accord with the findings for other DESs,[11,16,18,26,28,29,30] $\sigma_{dc}(T)$ of the present lithium-salt-doped systems can be well fitted by a modification of the empirical Vogel-Fulcher-Tammann (VFT) law[72,73,74,75] (lines in Figure 2a):

$$\sigma_{dc} = \sigma_0 \exp\left[\frac{-D_\sigma T_{VF\sigma}}{T - T_{VF\sigma}}\right] \quad (2)$$

Here $\sigma_0$ is a pre-exponential factor, $D_\sigma$ denotes the strength parameter, quantifying the deviations from Arrhenius temperature dependence[75] and $T_{VF\sigma}$ represents the Vogel-Fulcher temperature. Table 1 lists the obtained fit parameters. It should be noted that the viscosity of glass-forming liquids usually exhibits VFT temperature dependence[67] and the coupling of the ionic translational motions to the glassy freezing can be assumed to be the main reason for the validity of eq 2.

**Table 1.** Parameters obtained from the VFT fits of $\sigma_{dc}(T)$ and $\langle\tau\rangle(T)$ (eqs 2 and 3, respectively) of the investigated two glyceline/LiCl DESs (Figure 2). In addition, glass-transition temperatures derived from $\tau(T_g) = 100$ s and the fragility parameters calculated from $D_\tau$ are listed.

| | $T_g$ (K) | $T_{VF\sigma}$ (K) | $D_\sigma$ | $\sigma_0$ ($\Omega^{-1}$cm$^{-1}$) | $T_{VF\tau}$ (K) | $D_\tau$ | $\tau_0$ (s) | $m$ |
|---|---|---|---|---|---|---|---|---|
| glyceline + 1 mol% LiCl | 177 | 112 | 22.0 | 462 | 121 | 17.8 | 1.87×10$^{-15}$ | 49 |
| glyceline + 5 mol% LiCl | 181 | 116 | 20.4 | 230 | 123 | 18.6 | 1.30×10$^{-15}$ | 48 |

Figure 2b provides an Arrhenius plot of the mean dipolar α-relaxation times $\langle\tau\rangle$, as resulting from the fits of the permittivity spectra of the two investigated samples (closed symbols). The crosses show the corresponding data for pure glyceline.[28] Again, significant non-Arrhenius behavior is detected, characteristic of glass-forming liquids.[38,47,67] Figure 2b reveals an increase of the relaxation time upon the addition of lithium ions to glyceline, i.e., a slowing-down of the reorientational dipole motions. As the $\sigma_{dc}(1/T)$ and $\langle\tau\rangle(1/T)$ traces in Figure 2 appear as approximate mirror images, both quantities seem to be at least roughly coupled. It should be noted that recently perfect coupling of the two dynamics (ionic translational and dipolar reorientational) was reported for two DESs, including glyceline, and found to arise via the direct coupling of both quantities to the viscosity.[28] As discussed above, within this scenario the high ionic potential of the lithium ions in the present samples causes an enhancement of interactions between the different molecular and ionic constituents and, thus, an increase of viscosity upon LiCl addition. Just as the Stokes-Einstein and Nernst-Einstein relations predict a direct connection of the dc conductivity and the viscosity (discussed above), according to the Debye-Stokes-Einstein relation (predicting $1/\tau \propto D_r \propto T/\eta$, with $D_r$ the rotational diffusion coefficient) the reorientational relaxation time should also be directly related to the viscosity.[68,69,70,71] The increased viscosity due to LiCl addition then explains the increased relaxation time for higher salt content seen in Figure 2b. However, one should be aware that, in glass-forming systems, deviations from the above relations occur quite frequently[69,71,76,77] leading to certain decoupling effects which will be examined for the present DESs in section 3.3.

The strong influence of lithium addition on the dipolar dynamics becomes most obvious when comparing pure glycerol[78] (plusses in Figure 2b) with the mentioned glycerol/LiOTf system (triangles) as in this DES the lithium content is much larger than in the present ones (25 mol%).[16] Notably, the dipole reorientations in this DES are much slower than those in glycerol while those of the presently investigated DESs are faster, despite their lithium content. This is due to the generally shorter relaxation times of glyceline, compared to glycerol,[26] which become only slightly enhanced by the limited amount of added lithium ions. In Ref. 26, the faster dynamics in glyceline was ascribed to the partial breaking of the hydrogen network by the admixture of the rather large choline and chloride ions. Their ionic potential is smaller than for lithium and thus it is reasonable that the network is not strengthened by their admixture.

The $\langle\tau\rangle(T)$ data of the two investigated DESs can also be well fitted by a VFT law:[72,73,74,75]

$$\langle\tau\rangle = \tau_0 \exp\left[\frac{D_\tau T_{VF\tau}}{T - T_{VF\tau}}\right] \quad (3)$$

The obtained parameters are listed in Table 1, revealing similar, but not identical magnitudes of the strength parameters and Vogel-Fulcher temperatures describing the $\langle\tau\rangle(T)$ and $\sigma_{dc}(T)$ data. The deviations seem to indicate some minor but significant decoupling of the ionic and reorientational dynamics which will be treated in more detail below. To quantify the deviations of $\langle\tau\rangle(T)$ from Arrhenius behavior, usually the so-called fragility parameter $m$ is employed. It can be derived from $D_\tau$ using the relation $m \approx 16 + 590/D_\tau$.[79] The fragility of an ionically conducting glass former can considerably influence its dc conductivity, even at room temperature.[80] As documented in Table 1, for the two investigated DESs, we obtain very similar values of 49 (1 mol% LiCl) and 48 (5 mol%), similar to pure glyceline ($m = 47$).[26] Therefore, the differences in their conductivity values are not related to their fragilities.

In dipolar glass-forming liquids, the glass-transition temperature $T_g$ can be estimated applying the often-used criterion $\langle\tau\rangle(T_g) \approx 100$ s.[38] From an extrapolation of the VFT fit curves in Figure 2b, we deduce $T_g \approx 177$ and 181 K for the sample with 1 and 5 mol% LiCl, respectively (Table 1). From the DSC measurements, we estimate $T_g \approx 176$ and 178 K, respectively (dashed lines in Figure S1), which is in reasonable agreement with the dielectric results, bearing in mind the smeared-out character of the glass transition in the DSC traces. In Ref. 26 it was pointed out that the glass-transition temperature $T_g$ can substantially affect the dc conductivity of DESs and can even be relevant for its room-temperature value. The viscosity becomes large when approaching the glass transition. In the present case, the found increase of $T_g$ when adding LiCl therefore gives rise to a stronger increase of the viscosity upon cooling for the 5 mol% sample. Considering particles moving within a viscous medium and the corresponding Stokes-Einstein, Debye-Stokes-Einstein, and Nernst-Einstein relations, then rationalizes the detected slower reorientational dynamics and lower conductivity found especially at low temperatures for the 5% sample (Figure 2).

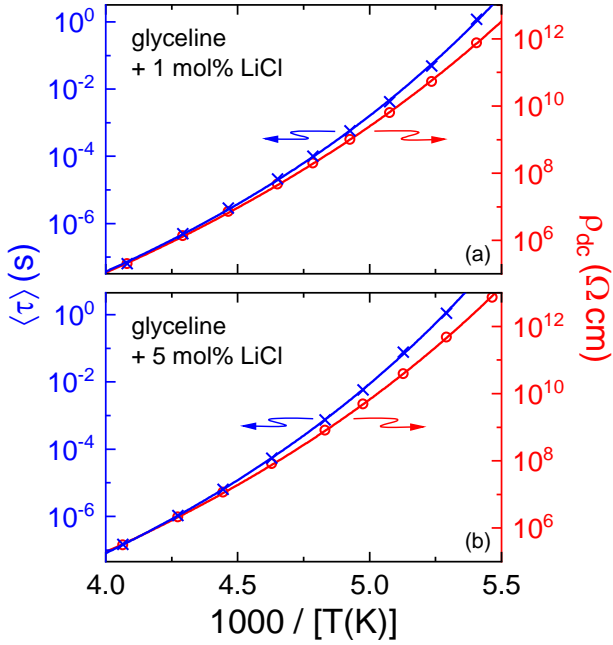

Figure 3. Arrhenius representation of the $\alpha$-relaxation times (crosses, left ordinates) and the dc resistivities (circles, right ordinates) of glyceline with 1 (a) and 5 mol% (b) LiCl. The left and right ordinates in each frame cover the same number of decades and their starting values are adapted to realize a match of both traces at the highest investigated temperature. The lines show the VFT fits from Figure 2 (eqs 2 and 3), using $\rho_{dc} = 1/\sigma_{dc}$ for the resistivity.

**3.3. Coupling of reorientational and translational dynamics.** For a visual check of the coupling of the dc ionic charge transport and the dipolar reorientations in the investigated samples, Figure 3 shows Arrhenius plots of $\langle\tau\rangle$ and the dc resistivity $\rho_{dc} = 1/\sigma_{dc}$ within the same frame.[81] Here, the left and right ordinates for $\langle\tau\rangle$ and $\rho_{dc}$, respectively, are adjusted to cover the same number of decades. Moreover, their starting values are chosen to achieve a match of both traces at high temperatures, since decoupling effects are usually less pronounced at elevated temperatures. As already concluded from Figure 2, the general temperature dependence of the two quantities is similar. However, with decreasing temperature deviations of up to a factor of five arise, evidencing some significant decoupling. This contrasts with the findings in pure glyceline, where nearly perfect coupling of the translational ionic and dipolar reorientational dynamics was found, i.e., $\langle\tau\rangle \propto \rho_{dc}$.[26,28] In contrast, reline exhibits qualitatively similar decoupling of $\langle\tau\rangle$ and $\rho_{dc}$ as observed in Figure 2.[28] Interestingly, in these two DESs and in ethaline (a mixture of ethylene glycol and ChCl) nearly perfect coupling of the reorientational motions with the viscosity was reported.[28] Making the reasonable assumption that this is also the case for the present glyceline samples with only few mol% Li-salt admixture, the $\langle\tau\rangle(T)$ data in Figure 3 can be supposed to show the same temperature dependence as the viscosity $\eta(T)$. Then the detected deviations of the $\langle\tau\rangle(T)$ and

$\rho_{dc}(T)$ curves, revealed in Figure 3 at low temperatures, imply that at least part of the ionic charge carriers in these DESs exhibit higher mobility than expected within the simple picture of translationally moving spheres within a viscous medium, the latter leading to $\rho_{dc} \propto \eta \propto \langle\tau\rangle$. As such deviations are absent in pure glyceline, this higher ionic mobility can be assigned to the small lithium ions, which find paths to diffuse faster than expected within the rather rigid structural network formed by the other constituents at low temperatures. As found in a recent NMR study,[82] in pure glyceline the charge transport is dominated at all temperatures by the choline cations rather than the smaller Cl anions. It seems plausible that the motions of these rather bulky ions comply with the above-mentioned simple scenario and are coupled to the viscosity (and, thus, to $\langle\tau\rangle$). NMR measurements of glyceline with added LiCl are desirable to confirm the above explanation of the detected decoupling effects.

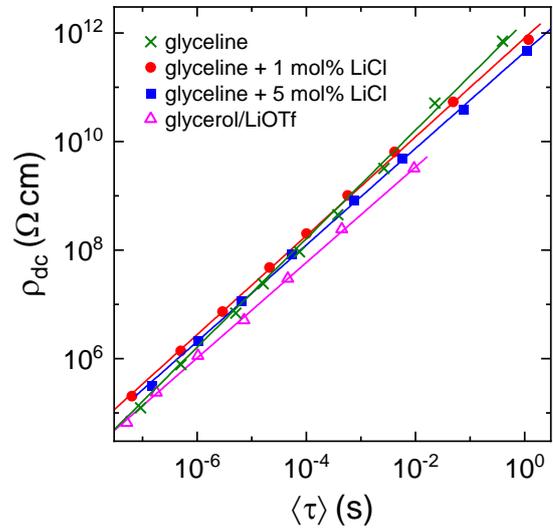

Figure 4. Dependence of the dc resistivity on the dipolar reorientational relaxation time for pure glyceline,[26] and for glyceline with 1 and 5 mol% LiCl investigated in the present work. For comparison, also data for a DES formed by glycerol and LiOTf are included.[16] The lines are linear fits with slope $\xi$, implying power-law behavior, $\rho_{dc} \propto \langle\tau\rangle^{\xi}$. For pure glyceline, a reasonable fit is obtained with $\xi = 1$, while $\xi$ is smaller than one for the other systems (1 mol% LiCl: $\xi = 0.91$, 5 mol% LiCl: $\xi = 0.89$, glycerol/LiOTf[16]: $\xi = 0.88$).

Finally, Figure 4 presents a double-logarithmic plot of the dc resistivity vs the $\alpha$-relaxation time as measured at different temperatures for the two investigated DESs. For comparison, also data for pure glyceline[26,28] and for the mentioned glycerol/LiOTf system[16] are included. The lines in Figure 4 are linear fits with slope $\xi$, which correspond to power laws, $\rho_{dc} \propto \langle\tau\rangle^{\xi}$, when considering the double-logarithmic axes. The data for pure glyceline can be reasonably fitted with $\xi = 1$.[26] This is in accord with the Debye-Stokes-Einstein relation $\sigma_{dc} \tau = $ const.[69,70,83] (Unfortunately, in literature the term "Debye-Stokes-Einstein relation" is inconsistently used and may also denote an equation relating $\tau$ and $\eta$ as mentioned above.[69]) In contrast, for the present glyceline samples with LiCl admixture, we find $\xi = 0.91$ (1 mol% LiCl) and $\xi = 0.89$ (5 mol%). This confirms the decoupling of the ionic translational and dipolar reorientational dynamics evidenced in Figure 3. Exponents $\xi < 1$ were previously also found for the glycerol/LiOTf system ($\xi = 0.88$) and other lithium-salt-based DESs.[16] Such behavior, corresponding to $\sigma_{dc}\,\tau^{\xi}$ = const., was previously also reported for various other glass-forming liquids and termed fractional Debye-Stokes-Einstein equation.[69,70,83]

Figure 4 reveals that, at low temperatures, for given relaxation times the LiCl addition to glyceline leads to an enhancement of the dc conductivity (i.e., smaller dc resistivity). If one considers the likely direct coupling of the reorientational dynamics and viscosity discussed above,[28] this again indicates enhanced low-temperature mobility of the small lithium ions due to decoupling from the viscosity. However, one should be aware that, for given *temperatures*, the lithium-doped samples exhibit reduced conductivity (see Figure 2a) due to their higher glass-transition temperatures and, thus, viscosities. Remarkably, without decoupling, these systems would reveal even lower $\sigma_{dc}$ at low temperatures.

## 4. CONCLUSIONS

In summary, we have performed detailed dielectric-spectroscopy measurements of the DES glyceline with two different admixtures of LiCl. Such DESs containing only few mol% of lithium salts are much more cost-efficient and sustainable that using a DES where the salt is one of the two main constituents.

The covered broad temperature range extends from the low-viscosity liquid state, above room temperature, to the deeply supercooled state at low temperatures, approaching the glass temperature. The obtained dielectric spectra provide clear evidence for an intrinsic dipolar relaxation process which can be assumed to be dominated by the reorientational motions of the glycerol molecules. The data are analyzed by applying an equivalent-circuit approach,[45] allowing the modeling of the complete spectra including the strong non-intrinsic electrode effects and providing unequivocal information on the intrinsic sample properties. Here, we especially focus on the dipolar relaxation time and on the dc conductivity, which should be high for possible electrolyte applications.

The slowing-down of the detected intrinsic relaxation process upon cooling follows pronounced non-Arrhenius behavior, clearly evidencing glassy freezing, which seems to be a common property of DESs. This is also in accord with the broadening of the detected relaxation features, signaling a distribution of relaxation times due to heterogeneity, and with the DSC results, revealing a typical glass-transition anomaly. The dc conductivity of the investigated DESs also exhibits non-Arrhenius temperature dependence, demonstrating that their translational ionic dynamics is also strongly influenced by the glassy freezing. With increasing LiCl content, the conductivity is reduced, and the dipolar dynamics slows down. This can be ascribed to a rise of the glass-transition temperature and the high ionic potential of the lithium ions, leading to enhanced interactions between the different DES constituents. These undesirable effects of lithium-salt addition, however, are much less pronounced than in a DES formed by glycerol and LiCl where the lithium salt is the sole HBA constituent.[16] Compared to this system, the conductivity in the present DESs is strongly enhanced due to the breaking of the hydrogen-bond network by the large choline ions having lower ionic potential than Li$^+$. We find that this can be traced back to a variation of the glass-transition temperature and, in this respect, these DESs with relatively small amounts of lithium behave significantly different than previously investigated systems with high lithium content.

A comparison of the deduced dc conductivity and reorientational relaxation times in the investigated DESs reveals significant decoupling of the corresponding translational ionic and reorientational dipolar dynamics, becoming increasingly pronounced at low temperatures. This makes it unlikely that a revolving-door-like mechanism, where the mobile ions move through paths opened by the rotation of the dipoles,[32,33,34] dominates the charge transport in the present systems. Just as recently reported for several other DESs,[16,26,29,30] the investigated samples exhibit a fractional Debye-Stokes-Einstein relation, $\sigma_{dc}\,\tau^{\xi}$ = const. ($\xi < 1$). In contrast, in pure glyceline $\xi = 1$ is found, signifying full coupling of charge transport and dipolar reorientations.[26] This implies that the small lithium ions are responsible for the observed decoupling in the present lithium-doped glyceline samples, most likely by exploring paths through the liquid structure that enable faster diffusion than expected for a viscous medium at low temperatures. While this conductivity enhancement is not dramatic in the present samples, it should be noted that such effects are the basis for enhanced conductivity, e.g., in superionic glasses.[67] Therefore, its further exploration aiming at the development of better DES-based electrolytes seems promising. Moreover, the present results again demonstrate the importance of the often-neglected glassy freezing of DESs and of decoupling effects for their dc conductivity. Obtaining additional microscopic information on the effect of lithium-salt addition, especially on its influence on the hydrogen-bond network and on cluster formation, e.g., using infrared spectroscopy and nuclear magnetic resonance measurements, is desirable.

## NOTES
The authors declare no competing financial interest.

## ACKNOWLEDGMENTS
This work was supported by the Deutsche Forschungsgemeinschaft (project no. 444797029). We thank M. Putz for performing the dielectric measurements.

# Supporting Information

for

# Ionic Conductivity of a Lithium-Doped Deep Eutectic Solvent: Glass Formation and Rotation-Translation Coupling

**A. Schulz, P. Lunkenheimer**\*, **and A. Loidl**

\* Corresponding author: peter.lunkenheimer@physik.uni-augsburg.de

Experimental Physics V, Center for Electronic Correlations and Magnetism, University of Augsburg, 86135 Augsburg, Germany

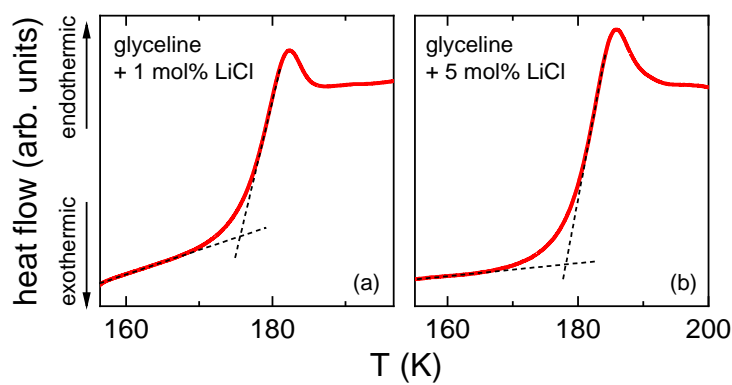

Figure S1. DSC heat flow for glyceline with 1 (a) and 5 mol% LiCl (b). The data were measured upon heating with 10 K/min after previous cooling with the same rate. The dashed lines indicate the estimation of the glass-transition temperatures leading to 176 and 178 K for the 1 and 5 mol% sample, respectively.